\documentclass[aps,pra,float,twocolumn,superscriptaddress]{revtex4}
\usepackage{graphicx}
\usepackage{braket}
\usepackage{amsmath}
\usepackage[colorlinks, citecolor=blue, urlcolor=red, linkcolor=blue]{hyperref}
\usepackage{amsfonts}
\usepackage{float}
\usepackage[font=small,format=plain,labelsep=period,labelfont=bf,textfont=normal,singlelinecheck=false,justification=raggedright]{caption}
\usepackage{xcolor}
\usepackage{comment}

\begin{document}

	\title{Quantum percolation in quasicrystals using continuous-time quantum walk}

	\author{Prateek Chawla}
	\affiliation{The Institute of Mathematical Sciences, C. I. T. Campus, Taramani, Chennai 600113, India}
	\affiliation{Homi Bhabha National Institute, Training School Complex, Anushakti Nagar, Mumbai 400094, India}
	\author{C. V. Ambarish}
	\affiliation{The Institute of Mathematical Sciences, C. I. T. Campus, Taramani, Chennai 600113, India}
	\affiliation{University of Wisconsin-Madison, Madison, WI 53706, USA}
	\author{C. M. Chandrashekar}
	\email{chandru@imsc.res.in}
	\affiliation{The Institute of Mathematical Sciences, C. I. T. Campus, Taramani, Chennai 600113, India}
	\affiliation{Homi Bhabha National Institute, Training School Complex, Anushakti Nagar, Mumbai 400094, India}



\begin{abstract}
	We study the percolation of a quantum particle on quasicrystal lattices and compare it with the square lattice. For our study, we have considered quasicrystal lattices modelled on the pentagonally symmetric Penrose tiling and the octagonally symmetric Ammann-Beenker tiling. The dynamics of the quantum particle are modelled using the continuous-time quantum walk (CTQW) formalism. We present a comparison of the behaviour of the CTQW on the two aperiodic quasicrystal lattices and the square lattice when all the vertices are connected and when disorder is introduced in the form of disconnections between the vertices. Unlike on a square lattice, we see a significant fraction of the quantum state localized around the origin in the quasicrystal lattices. With increase in disorder, the percolation probability of a particle on a quasicrystal lattice decreases significantly faster when compared to the square lattice. This study also sheds light on the fraction of disconnections allowed to see percolation of quantum particle on these quasicrystal lattices.
\end{abstract}	
	
	
\maketitle

\section{Introduction}
\label{sec:intro}
\noindent
The dynamics of particles in random media is described quite well by percolation theory\,\cite{K73, O86,SA94, BR06}. It is now a well established subject of research interest and has found applications across many disciplines of research\,\cite{SS94, KE09}. However, when the particle under consideration is a {\it quantum particle}, its dynamics are defined by quantum dynamics, and it has an added effect of quantum interference. A well-known consequence of interference effects  on quantum particle in disordered media on an infinite size lattice is Anderson localization\,\cite{A58,LR85,EM08}, a phenomenon where the interference of different phases picked up by the quantum particle travelling along various routes in a random or disordered system leads to strong localization of the particle's wavefunction around the origin.  This effect has also been experimentally observed in some disordered systems\,\cite{AALR79,SBFS07,CJ08,CA13}. However, on a finite sized system with a very small amount of disorder, the spread of the tail of the wave packet allows for some small probability of finding the quantum particle at some distance beyond the origin\,\cite{CB14, CMB14}. This makes quantum percolation a more interesting field of research leading to behaviours in the dynamics that are not seen in its classical counterpart\,\cite{KE72,SAB82, VW92, MDS95, SF09, QPLN09}.

Quantum walk\,\cite{GVR58,RPF86,ADZ93,DAM96, FG98}, a quantum mechanical analogue of the classical random walk is one of the efficient ways to model the controlled dynamics of a quantum particle. It has also been efficiently used to develop many quantum algorithms\,\cite{JK03, IKS05, YKE08, V12, ANAV17} and schemes for quantum simulations\,\cite{ SF92, MRL08, KRBD10, C11, C13, MMC17}. The quantum walk has been described in two prominent forms, the continuous-time quantum walk (CTQW) and discrete-time quantum walk (DTQW). Quantum interference is a dominating factor in dynamics of quantum walks and by introducing disorder into the dynamics, localization of quantum states has been demonstrated\,\cite{J12, C12, COB15}. Therefore, quantum walk is one of the most suitable ways to study the dynamics of quantum particle on various lattices. In this work we use CTQW to describe the dynamics of quantum particle on quasicrystal lattice and study quantum percolation. 

Quasicrystals do not show periodicity and have no long range translational symmetry. Yet they present long-range order and have a well-defined Bragg diffraction pattern\,\cite{SBG84, CJ94,S95, S04, B09, BG13, S18}. The diffraction patterns from quasicrystal lattices reveal fivefold, eightfold and tenfold symmetries that cannot originate from a periodic arrangement of unit cells. It has been demonstrated that quasicrystal lattices may be seen to originate from a projection of a periodic lattice in a higher dimension \cite{P74, KLZ12, KRZ13, JD13}. In this paper, we consider lattices based on fivefold (pentagonally) symmetric Penrose tiling\,\cite{P74} and eightfold (octagonally) symmetric Ammann-Beenker tiling\,\cite{GS86}. These tilings are the basis for designing two-dimensional quasicrystals with self-similarity in both real and reciprocal spaces. It is more convenient to explore the dynamics of a quantum particle on quasicrystals with vertices of different degrees via CTQW rather than DTQW, as the latter would require different dimensional coin spaces at each vertex. Thus, using CTQW we study quantum percolation on two quasicrystal lattices and compare the results with the dynamics seen on square lattice. We show that for a square lattice the wave packet spreads away from the origin when compared to the lattices based on Penrose and Ammann-Beenker tilings, where a significant fraction of the wave packet remains localised around the origin. With increase in disorder in the form of disconnections between the vertices in the lattice  (irrespective of the initial position), the percolation probability of a particle on both quasicrystal lattices under consideration decreases significantly faster when compared to the square lattice. 
This study also sheds light on the effect of disorder on the dynamics of a quantum particle on a quasicrystal lattice. 

This paper is organized as follows. In section\,\ref{sec:ctqw}, we define a CTQW on a graph, and state the conventions used throughout the paper. Section\,\ref{sec:res} contains the results of our numerical simulations of quantum percolation on the square lattice and aperiodic lattices, along with comparisons between the results. We conclude with summary of our observations in section\,\ref{sec:conc}.


\begin{table*}


\begin{minipage}[c]{\textwidth}
~
	\centering
	\begin{tabular}{ccc}
		\includegraphics[width=0.3\textwidth]{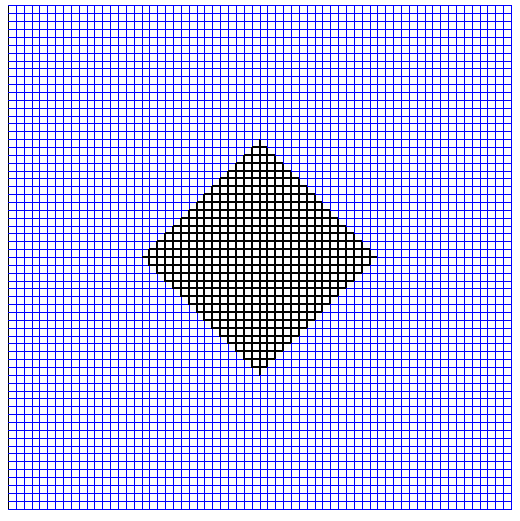} &
		\includegraphics[width=0.3\textwidth]{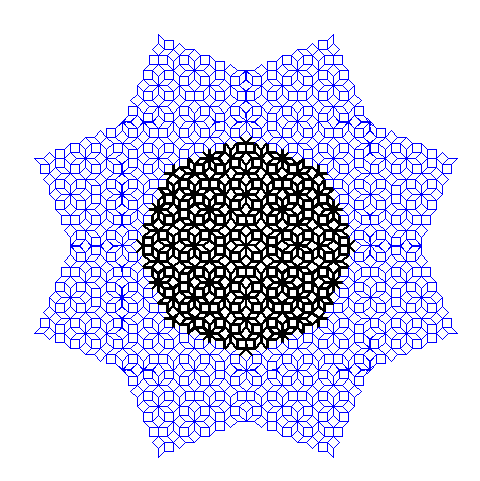} &
		\includegraphics[width=0.3\textwidth]{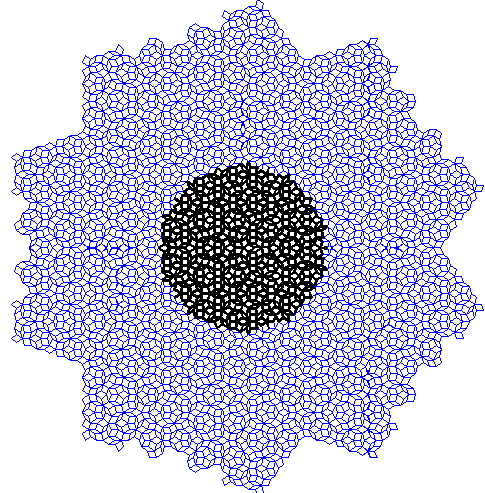} \\	
		(a) & (b) & (c)
	\end{tabular}
	\captionof{figure}{Illustrations of the lattices considered in this paper. Fig. (a) shows a lattice based on a 65 x 65 square tiling, (b) shows a quasicrystal lattice based on a three iteration Ammann-Beenker tiling, and (c) illustrates another quasicrystal lattice based on a four-iteration Penrose tiling. Each lattice has a 15 hopping length percolation test zone marked about the center.
	\label{fig:tilings}}
\end{minipage}
\begin{minipage}[c]{\textwidth}
	\centering
	\begin{tabular}{cc}
		\includegraphics[width=0.45\textwidth]{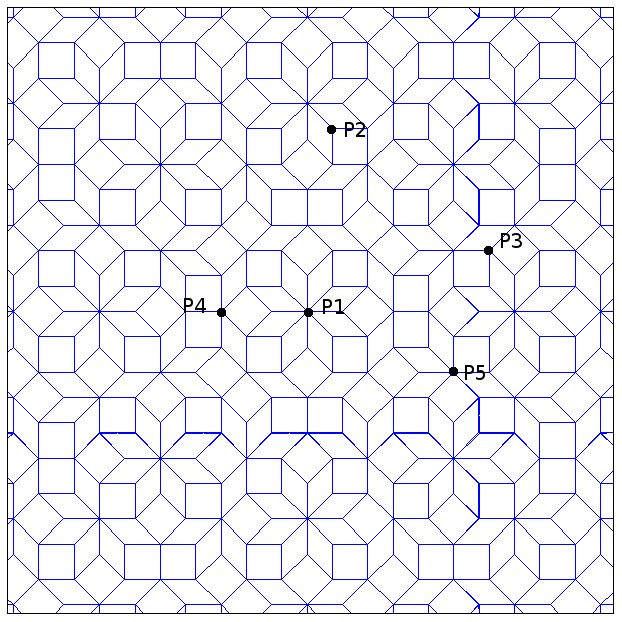} ~~&~~ 
		\includegraphics[width=0.45\textwidth]{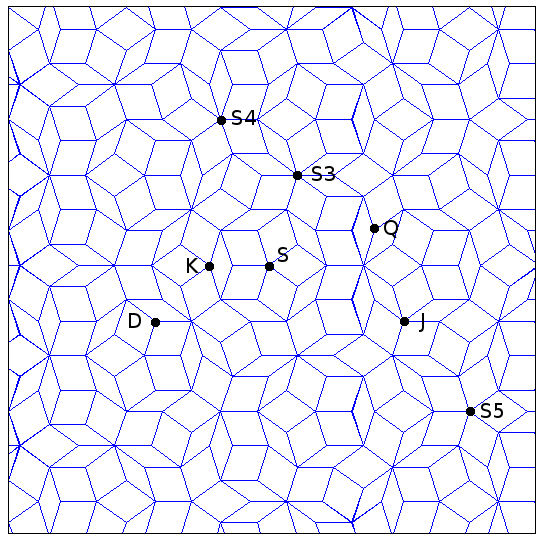} \\
		(a) & (b)
	\end{tabular}
	\captionof{figure}{An illustration of the different kinds of points on the lattices considered. Fig. (a) shows the different possible initial points on the lattice based on Ammann-Beenker tiling, and (b) shows the different possible initial points on the lattice based on Penrose tiling.
	\label{fig:latticeMarked}}

\end{minipage}
\end{table*}

\section{Continuous-time quantum walk}
\label{sec:ctqw}
\noindent
Consider an undirected graph $\Gamma(V,E)$, where $V$ is a set of $N$ vertices and $E$ are the edges. Let $A$ be the adjacency matrix of $\Gamma$, defined as
\begin{equation}
	\label{eq:eq2.1}
	A_{ij} := \begin{cases}
		1 & \text{edge } (i,j) \in E \\
		0 & \text{otherwise}
	\end{cases}. 
\end{equation}
The resulting matrix $A$ is a real-valued matrix, symmetric about the main diagonal. The vertices are labelled by the computational basis states $\{\ket{1}, \ket{2}, ..., \ket{N}\}$, and the quantum state of the entire graph at any time $t$ is represented by a normalized vector $\ket{\psi(t)}$, defined to be
\begin{equation}
	\label{eq:eq2.2}
	\ket{\psi(t)} = \sum_{l=1}^{N} \alpha_l\ket{l} \;\;\;\; \alpha_l \in \mathbb{C}.
\end{equation}
While classical Markovian processes obey the master equation, the CTQW obeys the Schr\"{o}dinger equation, and the state $\ket{\psi(t)}$ transforms with time as,
\begin{equation}
	\label{eq:eq2.3}
	i\hbar \frac{\partial}{\partial t} \ket{\psi(t)} = H_\Gamma \ket{\psi(t)}.
\end{equation}
where $H_\Gamma$ is the Hamiltonian matrix, defined for the graph as 
\begin{equation}
	\label{eq:eq2.4}
	\begin{aligned}
		H  &= \gamma L = \gamma(D-A), \\
		\implies H_{ij} &= \begin{cases}
			-\gamma & i\neq j,\;\; (i,j) \in E \\
			0 & i\neq j,\;\; (i,j) \notin E \\
			d_{ii}\gamma & i = j
		\end{cases} 
	\end{aligned}.	
\end{equation}

In the preceeding expression, $D$ is a diagonal matrix with the diagonal element $d_{ii}$ being the degree of vertex $i$, $\gamma$ is a finite constant characterizing the transition rate on the graph $\Gamma$, and $A$ is the adjacency matrix. For the purposes of this work, $\gamma$ has been taken to be $1$. 

The Hamiltonian is time-independent, and thus the formal solution to Eq.\,\eqref{eq:eq2.3} is given by the time evolution operator $U$, defined as,
\begin{equation}
	\label{eq:eq2.5}
	U = e^{-iH_\Gamma t},
\end{equation}
\noindent
in the units of $\hbar$. Since the Hamiltonian is real and symmetric (and hence Hermitian), $U$ is necessarily unitary, and thus the norm of $\ket{\psi(t)}$ is conserved, as is required for a CTQW.

The initial state of the particle is completely localized at one particular vertex. We define a subset of the vertices around the initial vertex that may be reached in a certain number of steps. We use CTQW to study the dynamics of percolation of this particle. To simulate the effect of disorder on the system, we randomly remove a certain fraction of the edges, which we term as {\it edge disconnection fraction}. We then let the particle perform a CTQW as usual on this graph. The probabilities from a number of runs are averaged to get a general estimate of how the percolation changes by changing the amount of disorder.

For the purposes of this paper, we have considered a hopping zone of length $40$, and the particle is considered to have percolated if $\geq 2\%$ of its total probability lies outside the hopping zone. 


\section{Quantum percolation using CTQW}
\label{sec:res}


Fig.\,~\ref{fig:tilings} shows the illustration of the three lattices we have considered in this paper. The representations built here are so that the viewing is more convenient. The actual calculations have been done on a 41-iteration Ammann-Beenker tiling and a 7-iteration Penrose tiling. The {Ammann-Beenker tiling is octagonally symmetric}, with square and rhombuses as its basic elements. There are different kinds of points in the Ammann-Beenker tiling, which are divided into categories as shown in Fig.\,\ref{fig:latticeMarked}~(a). The pentagonally symmetric Penrose tiling is formed by using two fat and thin rhombi as the basic elements. The different kinds of points in the Penrose {tiling} are shown in Fig.\,\ref{fig:latticeMarked}~(b).

\subsection{Probability distribution of the CTQW}

In this subsection, we take a look at the probability distribution of the particle whose evolution is described by the CTQW on the three types of lattices considered.
\begin{figure}[!ht]
	\centering
	\includegraphics[width=\linewidth]{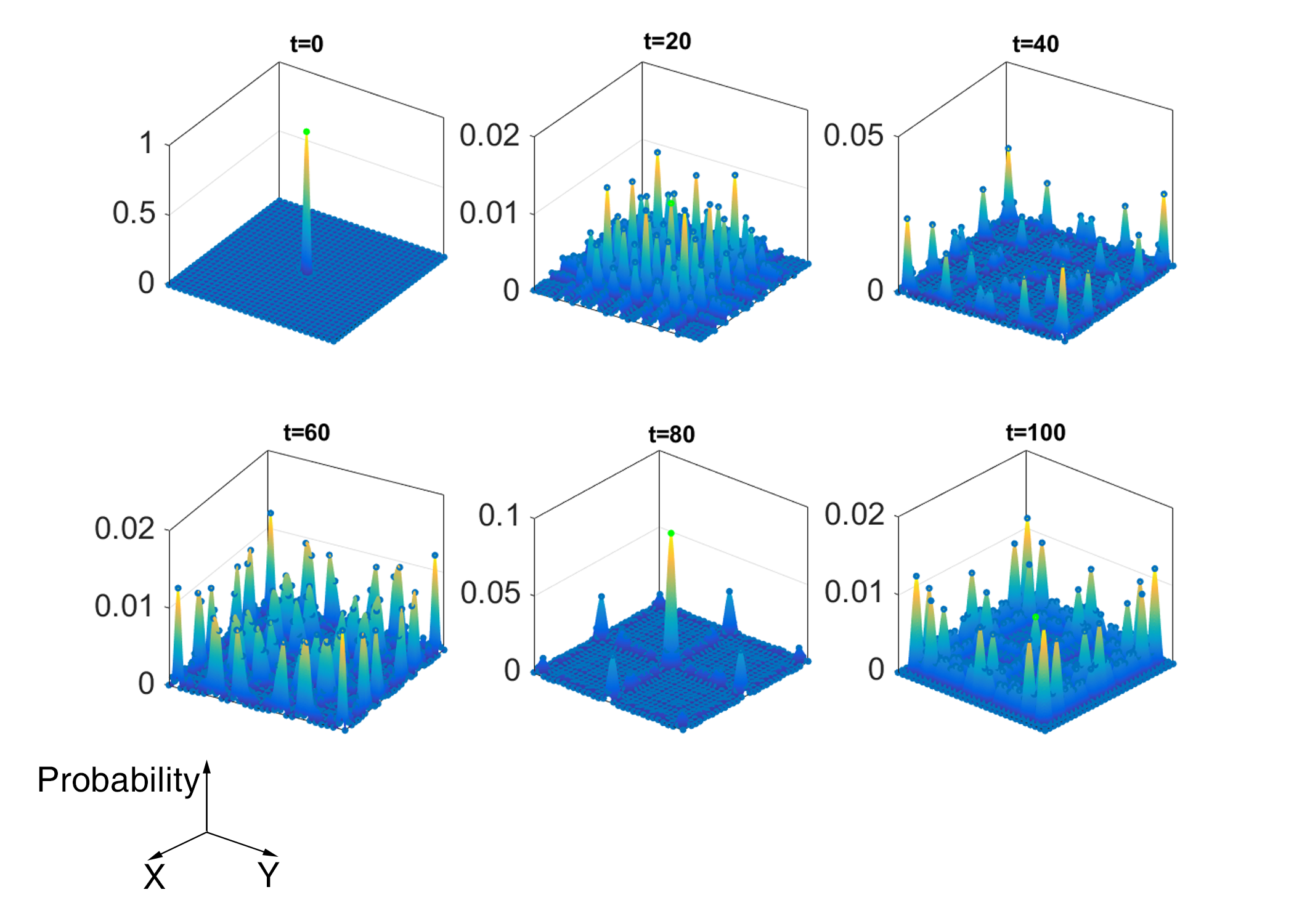}
	\caption{Evolution of the probability distribution on a 30x30 square tiling with time by CTQW of a point initially in the center.
		\label{fig:rect30}}
\end{figure}
\begin{figure}[!ht]
	\centering
	\includegraphics[width=\linewidth]{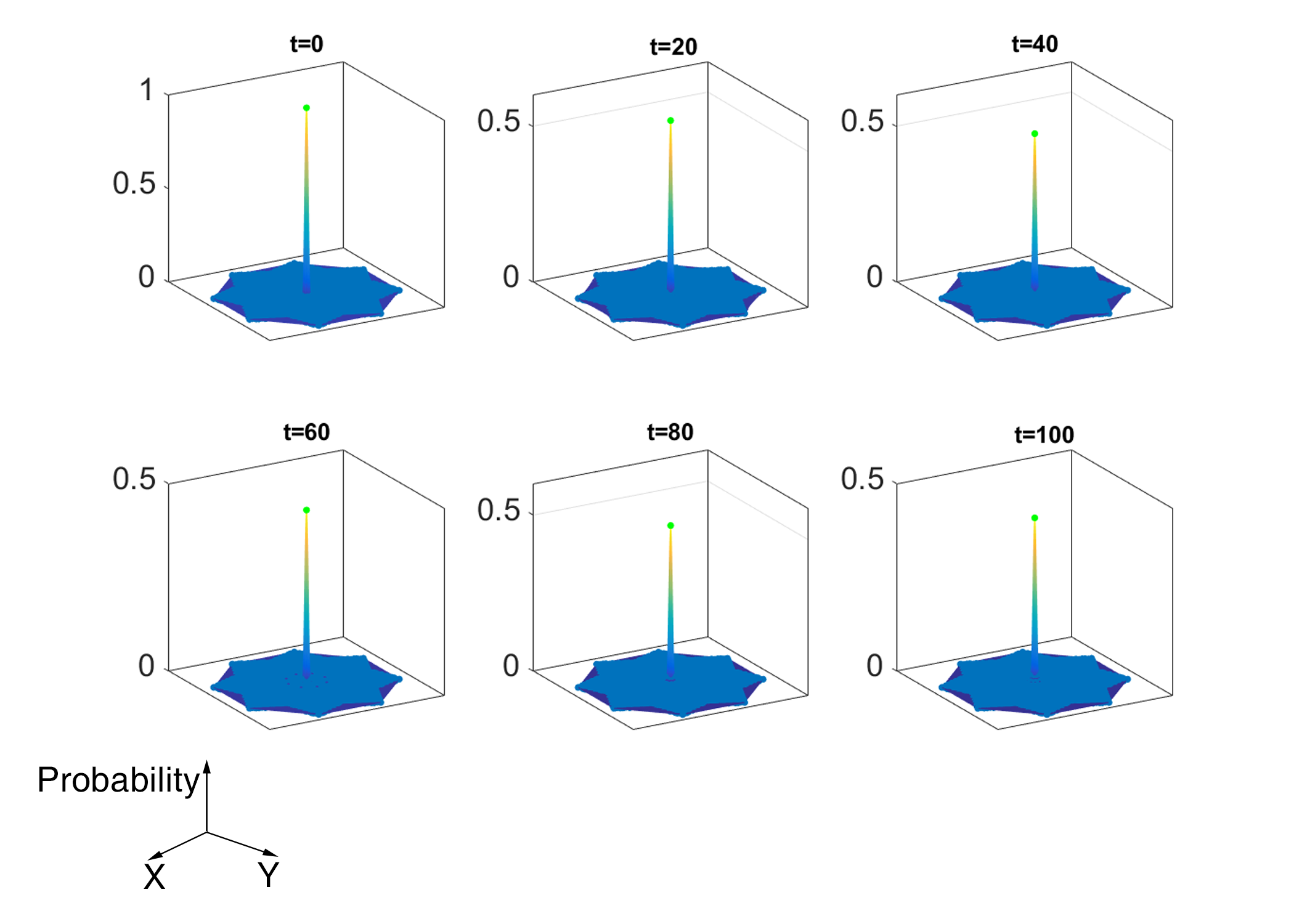}
	\caption{Evolution of the probability distribution on a three iteration Ammann-Beenker tiling with time by CTQW of a point initially in the center.
		\label{fig:ab3}}
\end{figure}
\begin{figure}[!ht]
	\centering
	\includegraphics[width=\linewidth]{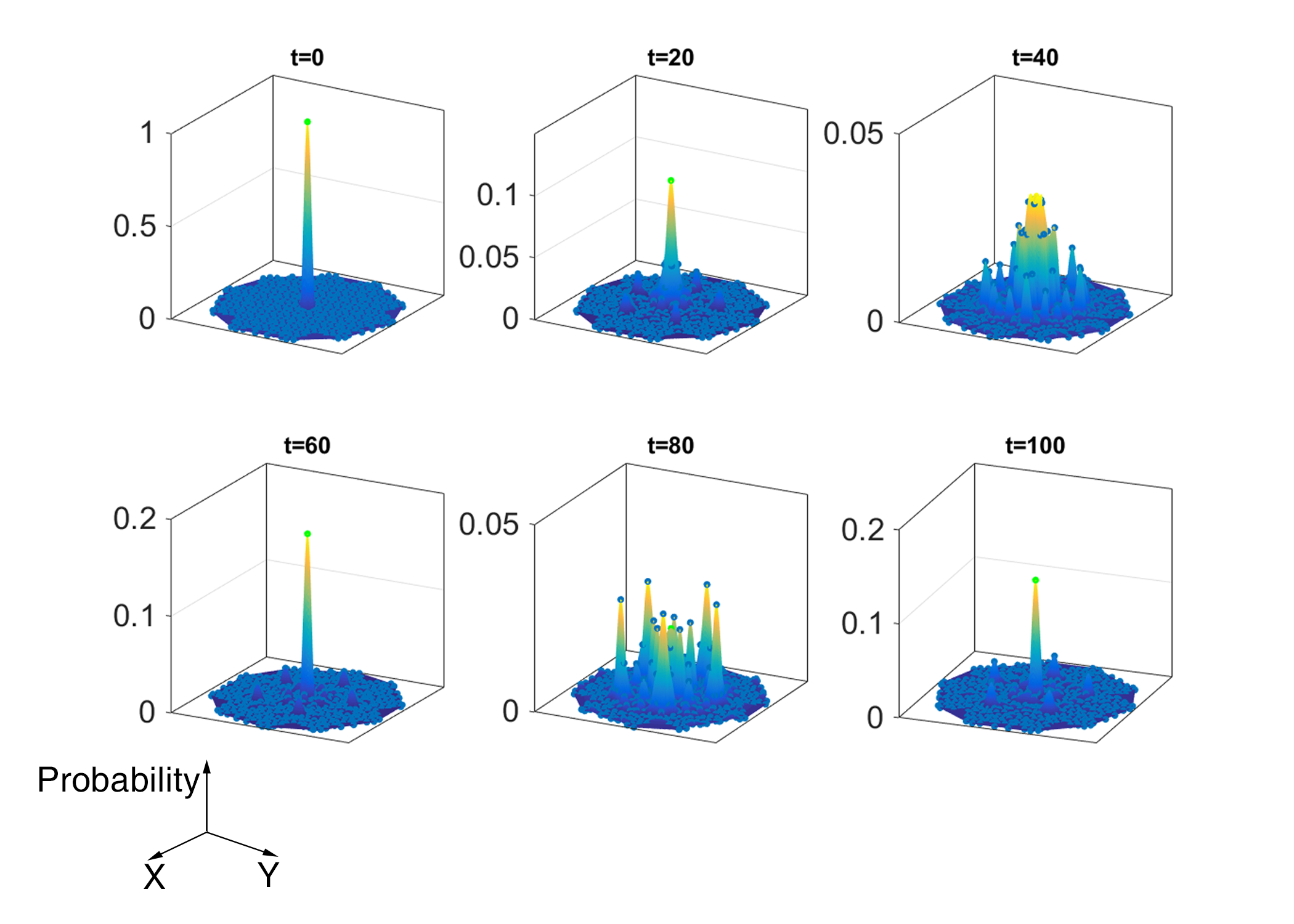}
	\caption{Evolution of the probability distribution on a four-iteration Penrose tiling with time by CTQW of a point initially in the center.
		\label{fig:penrose4}}
\end{figure}

From Figs.\,\ref{fig:rect30}, \ref{fig:ab3} and \ref{fig:penrose4}, we observe that the probability amplitude spreads much faster on a square lattice than in an Ammann-Beenker or a Penrose lattice. On both the quasicrystal lattices we see a significant fraction of amplitude being around the origin even after evolving for some time $t$. Among the two quasicrystal lattice, signature of being wavepacket of the particle being localized around the origin is more prominent in Ammann-Beenker lattice (Fig.\,\ref{fig:ab3}) when compared to the Penrose lattice (Fig.\,\ref{fig:penrose4}). Choosing different points in the quasicrystal lattice as initial position of the walker will alter the probability distribution, however, the overall behaviour will not see a significant change. This becomes more evident when we study the percolation probability when we have no disorder or disconnections in the lattice structure in the following subsection.

\subsection{Evolution of percolation probability with time}

\begin{figure}[!ht]
	\centering
	\includegraphics[width=\linewidth]{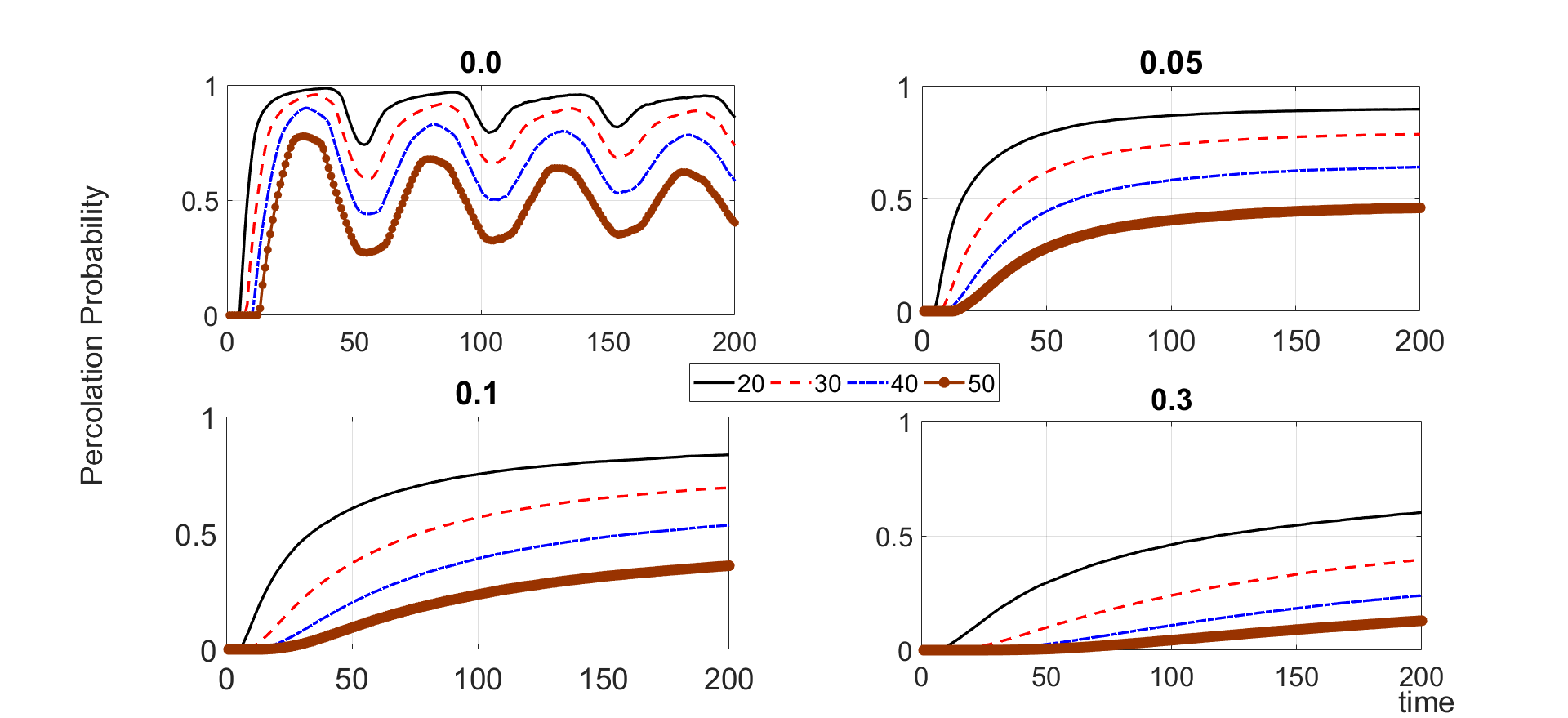}
	\caption{Probability of finding the particle outside the test zone in a 100x100 square tiling where the different test zones are defined based on hop-lengths from center. The test zones are 20, 30, 40, and 50 hop lengths as shown in the legend. The plots are produced for different amounts of disconnections. Going clockwise from the top left, they represent 0\%, 5\%, 10\%, and 30\% disconnections in edges.
		\label{fig:rect_time2}}
\end{figure}

The spread of the probability distribution with time on a finite-sized lattice is studied by considering the lattice in the form illustrated in Fig.\,\ref{fig:tilings}. The probability of finding $\geq 2\%$ of the probability distribution of the particle outside the hopping zone is considered as percolation probability. On a well-connected lattice without any disconnections between the vertices, at the edge of the hopping zone we will see and an effect of quantum state hopping in and out of the hopping zone in the form of oscillations in the percolation probability. However, with time the propogating part of wavepacket would move away from the hopping zone and the oscillation should dimnish with time. This is prominent for a square lattice as shown in Fig.\,\ref{fig:rect_time2}. This result can be derived analytically as well, and the states seen are found to be Bloch states. With an increase in amount of disorder in form of disconnections between the vertices in the lattice, we see a decrease in percolation probability showing the signature of a fraction of the wavepacket being trapped around the disconnected vertices. For plots with disconnections, we do not see any oscillation in probability distribution and this is most likely due to effect of averaging, as we have averaged the probabilities over 50 runs. Even though the edge disconnection fraction is fixed, the positon of dislocations may be different during each run. For a very small amount of disconnections, the probability of being found in a particular region reaches an equilibrium value faster with time $t$. For a lattice with a high amount of disconnections, the time taken to saturate to an equilibrium value is very high, and effectively, the probability of percolation does not saturate in a practical time scale.

\begin{figure}[!ht]
	\centering
	\includegraphics[width=\linewidth]{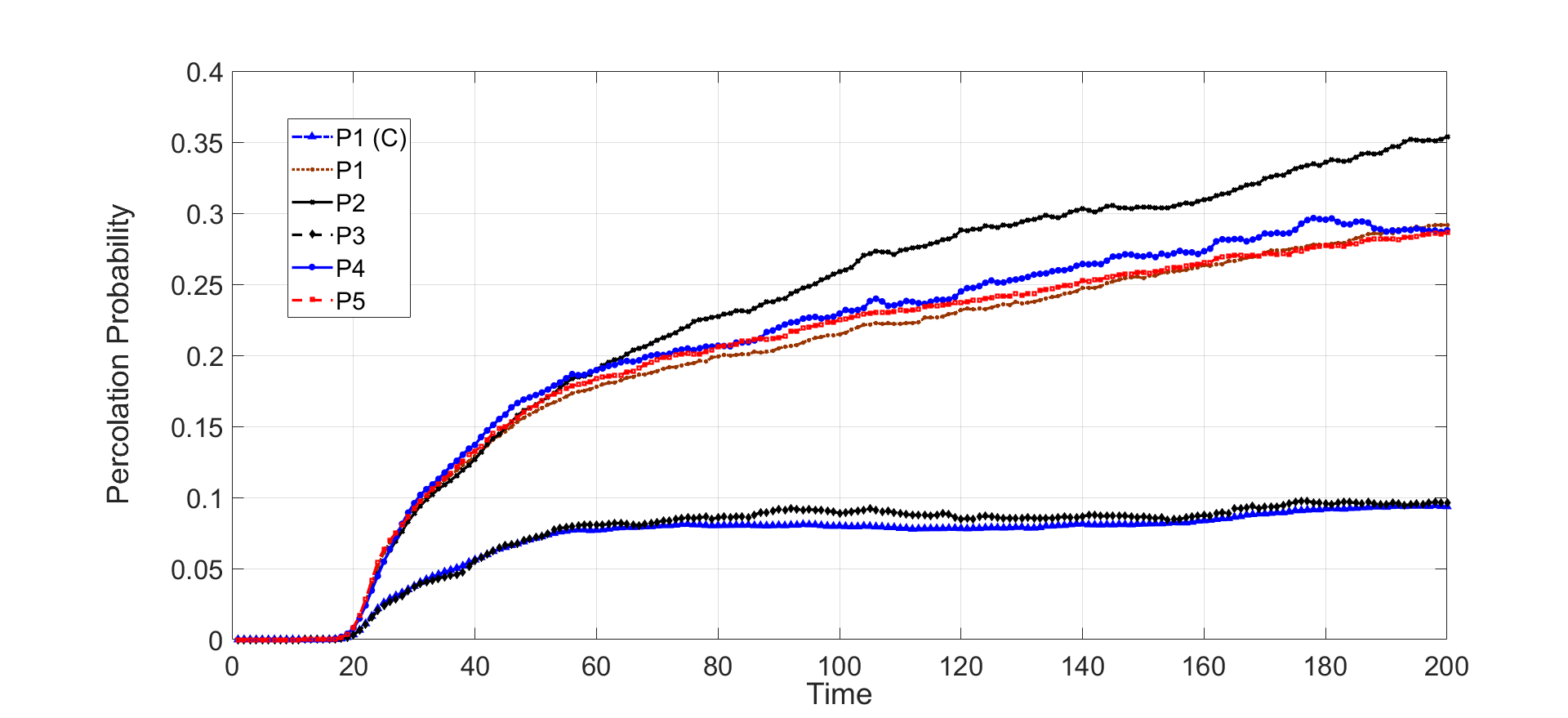}
	\caption{Percolation probability with time for different points on the Ammann-Beenker tiling with no disconnections. The curves are labeled for the different initial points defined in Fig.\,\ref{fig:latticeMarked}~(a) 
		\label{fig:AB_time}}
\end{figure}

\begin{figure}[!ht]
	\centering
	\includegraphics[width=\linewidth]{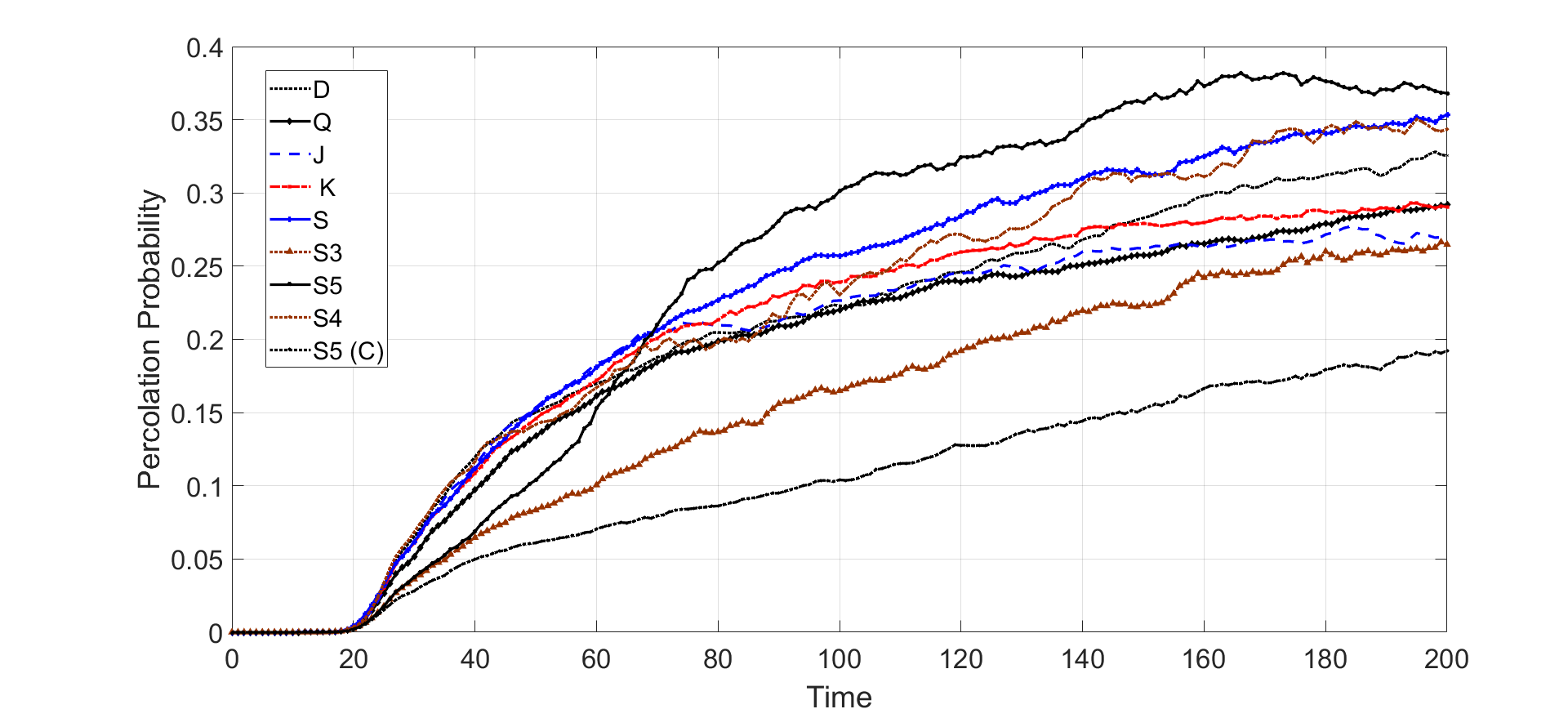}
	\caption{Percolation probability with time for different points on the Penrose tiling with no disconnections. The labels of the different curves are defined for different possible initial points as seen in Fig.\,\ref{fig:latticeMarked}~(b).
		\label{fig:penrose615v1_time}}
\end{figure}

The initial position of the particle on a quasicrystal lattice plays a noticable role on the percolation probability of the particle. In Fig.~\ref{fig:AB_time} and Fig.\,\ref{fig:penrose615v1_time}, we show the evolution of percolation probability with time for different starting points on a well connected Ammann-Beenker and Penrose tiling, respectively.  
One can notice that even without any disconnection we can see a significant difference in the percolation probability of the particle depending on the starting point of the particle. Overall a general behaviour indicates a better percolation probability on a Penrose tiling when compared with the Ammann-Beenker tiling.

\begin{figure}[!ht]
	\centering
	\includegraphics[width=\linewidth]{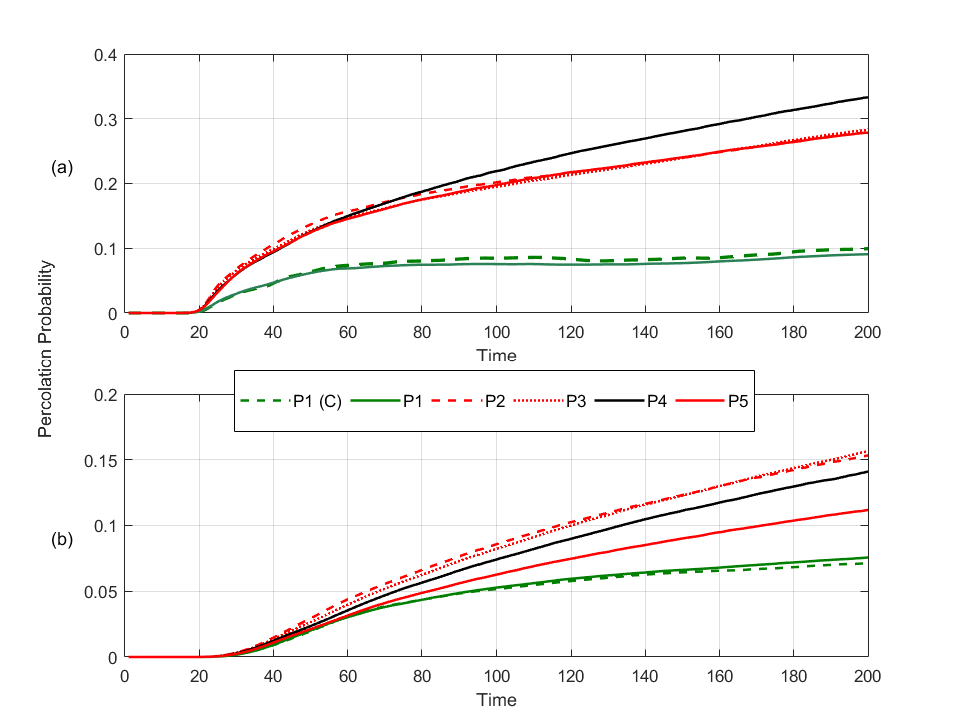}
	\caption{Percolation probability with time for different points on the Ammann-Beenker tiling with (a) 1\% disconnections (b) 10\% disconnections in a test zone of hopping distance 40. The initial points are labelled as marked in Fig.\,\ref{fig:latticeMarked}~(a)
		\label{fig:ABTimeWithDisconnections}}
\end{figure}

\begin{figure}[!ht]
	\centering
	\includegraphics[width=\linewidth]{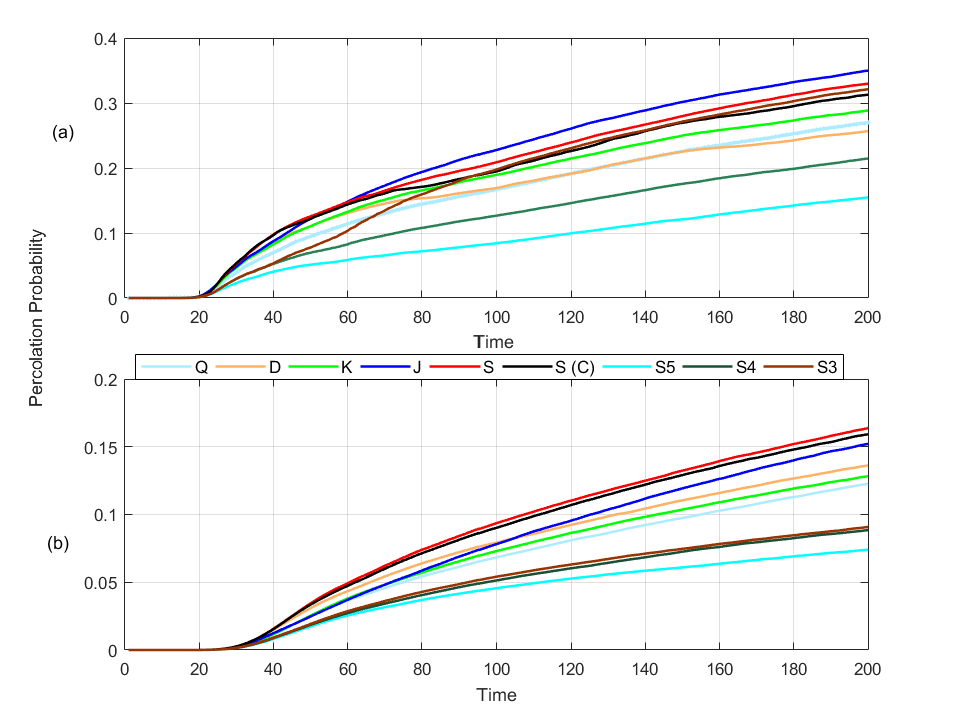}
	\caption{Percolation probability with time for different points on the Penrose tiling with (a) 1\%, and (b) 10\% edge disconnections in a test zone of hopping distance 40. The initial points are labelled as shown in Fig.\,\ref{fig:latticeMarked}~(b)
		\label{fig:penrose615v1_comb}}
\end{figure}

In Fig.~\ref{fig:ABTimeWithDisconnections} and Fig.\,\ref{fig:penrose615v1_comb} we show the percolation probability with time on an Ammann-Beenker and Penrose tiling, respectively, with an edge disconnection fraction of 1\% and 10\% between the vertices.  The effect of starting point of the particle reflects exactly the same way as does for dynamics without any disconnection, an overall decrease in percolation probability with increase in edge disconnection fraction is clearly evident here. 
 


\subsection{Evolution of percolation probability with edge disconnection fraction}

\begin{figure}[!ht]
	\centering
	\includegraphics[width=\linewidth]{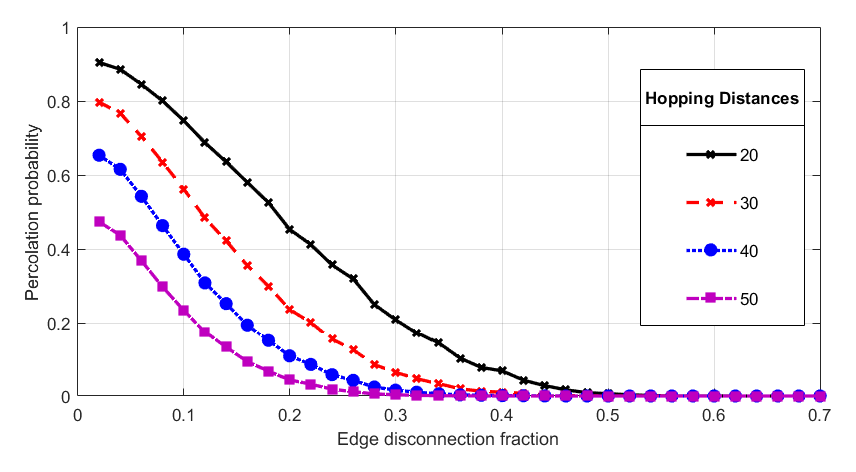}
	\caption{Plot of percolation probability with fraction of edge disconnections in a 100x100 square tiling for different hopping distance after time 200. Percolation probability is defined to be the probability of the particle being found outside the test zone (hopping distance).
		\label{fig:rect_prob}}
\end{figure}

To understand the effect of edge disconnections on the percolation probability we consider the effect of edge disconnections on the percolation probability. This will allow us to identify the fraction of disconnections allowed to see the percolation of quantum state on these lattices. In Fig.\,\ref{fig:rect_prob} we show the variation of percolation proability with an increase in edge disconnection fraction on a square lattice with different hopping distances. The probability of percolation vanishes after the fraction of disconnections exceed 50\% for all considered test zones. However, with increase in hopping distance, only a lower fraction of edge disconnection allows percolation of quantum states.

\begin{figure}[!ht]
	\centering
	\includegraphics[width=\linewidth]{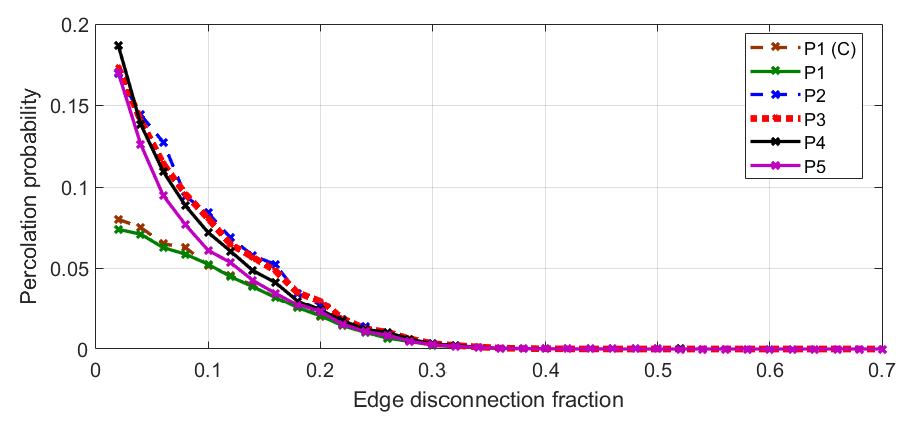}
	\caption{Plot of percolation probability with fraction of edge disconnections in a Ammann-Beenker tiling in a test zone of hopping distance 40. The initial points taken are as mentioned in Fig.\,\ref{fig:latticeMarked}~(a).
		\label{fig:ABEdgeProb}}
\end{figure}

\begin{figure}[!ht]
	\centering
	\includegraphics[width=\linewidth]{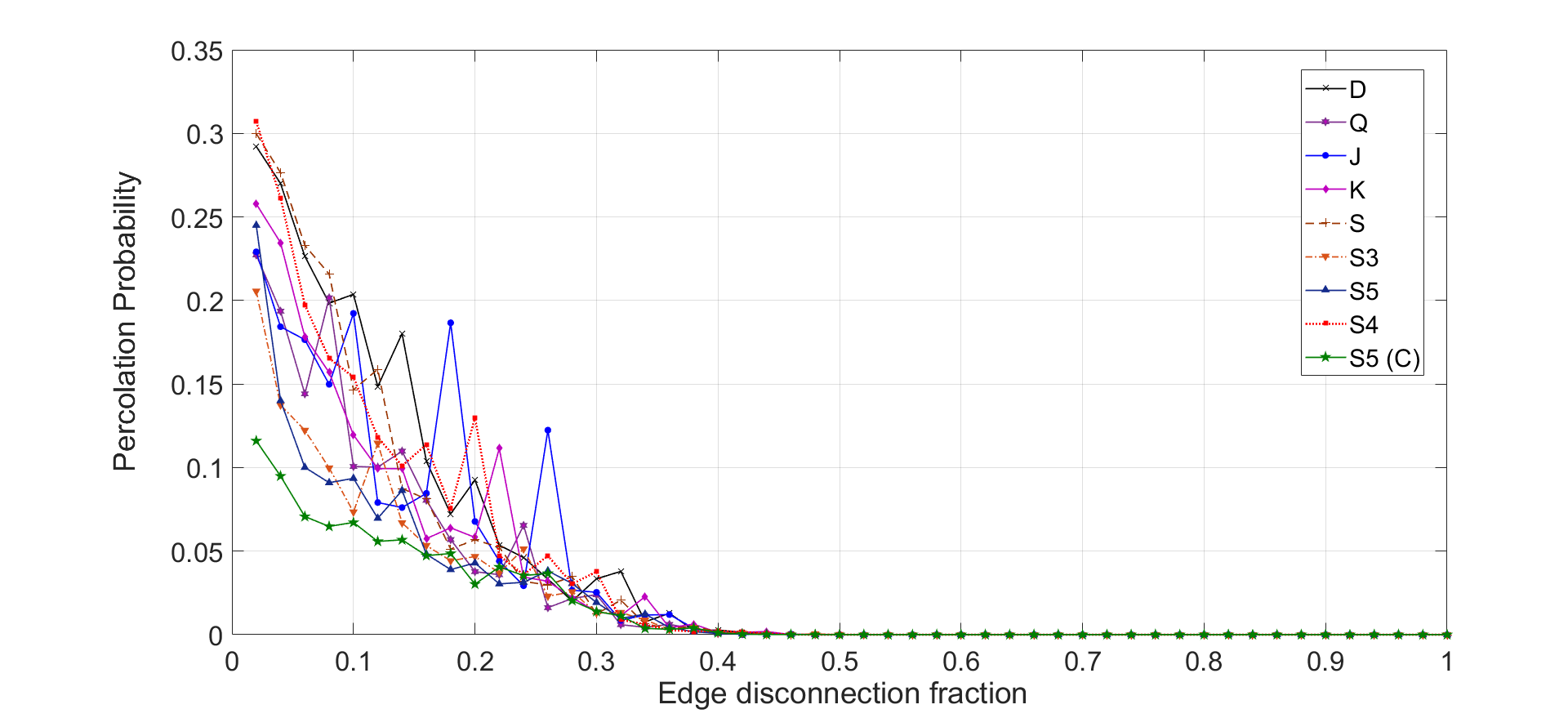}
	\caption{Plot of percolation probability with edge edge disconnection fraction in a Penrose tiling for various initial points in a zone of hopping distance 40. The initial points taken are as mentioned in Fig.\,\ref{fig:latticeMarked}~(b)
		\label{fig:penrose615v2_edgeProb}}
\end{figure}

In Figs.\,\ref{fig:ABEdgeProb} and \ref{fig:penrose615v2_edgeProb} we show the percolation probability for a particle on Ammann-Beenker and Penrose lattices as a function of the edge disconnection fraction. The percolation probability depends on the choice of the starting point, however, percolation probability vanishes after 32\% and about 40\% of edge disconnection fraction is introduced in Amman-Beeker and Penrose lattices, respectively.

\par 


In Fig.\,\ref{fig:latticesCombined} (see appendix) we show a comparison of the results obtained by simulating the CTQW on different lattices with various starting points. We have considered the percolation probability with time for different values of edge disconnection fraction and the behaviour of probability of percolation with edge disconnection fraction for time 200. On comparison we can say that the percolation probability on square lattice is always higher that the two quasicrystal lattices and percolation probability on penrose tiling is higher than percolation probability on Amman-Beenker lattice.

\section{Conclusion}
\label{sec:conc}
In this work, we have simulated the behaviour of a particle performing a CTQW on lattices modelled on Penrose and Ammann-Beenker tilings. We have then compared the results to those seen on a square lattice. It is clearly seen that due to the aperiodic nature of the Penrose and Ammann-Beenker tilings, the selection of the initial point plays a significant role in the percolation behaviour of the particle. 

In order to study the dynamics of percolation on different lattices, we have numerically evaluated the percolation probability with time for different values of edge disconnection fraction, and as a function of edge disconnection fraction as well. It is seen that the probability of a particle to percolate out of a test zone on a square lattice shows a pattern of oscillations with reducing amplitudes with time, as expected. The probability for a square lattice also seems to saturate when a small amount of disorder is introduced. 

For the case when the walk takes place on an aperiodic tiling, the particle shows a significantly lower probability to percolate out of the test zone. The Penrose tiling shows a higher probability of percolation compared to the Ammann-Beenker tiling, but the difference is very slight and reduces as the amount of disorder increases. Compared to the square tiling, for all possible initial positions on the aperiodic tilings, the probability of percolation reduces faster as the amount of disorder increases.

It can therefore be concluded that quasicrystal lattices show a significantly less percolation probability, and hence a slower speed of percolation of a quantum particle as compared to square lattices, and thus can be potentially used to store a quantum states for a higher amount of time. This has the potential application in creating lattices that may show higher coherence times than rectangular lattices, regardless of the amount of disorder.

\vskip 0.2in
\noindent
{\bf Acknowledgment:} CMC would like to thank Department of Science and Technology, Government of India for the Ramanujan Fellowship grant No.:SB/S2/RJN-192/2014. We also acknowledge the support from Interdisciplinary Cyber  Physical Systems(ICPS) programme of the Department  of  Science and Technology, India, GrantNo.:DST/ICPS/QuST/Theme-1/2019/1.


\clearpage

\appendix

\section{}

\renewcommand\thefigure{\thesection\arabic{figure}} 
\setcounter{figure}{0}   

In this section, we have collated results from the figures shown in the rest of the paper to make it easy to make a direct comparison between the values for different lattices. From the Fig.\,\ref{fig:latticesCombined}, it is seen that percolation on the quasicrystal lattice takes place slower than on a square lattice, and that this effect is prominent even with higher amounts of disorder.

\begin{table}[H]
	\begin{minipage}[c]{\textwidth}
		\centering
		\begin{tabular}{cc}
			\includegraphics[width=0.5\linewidth]{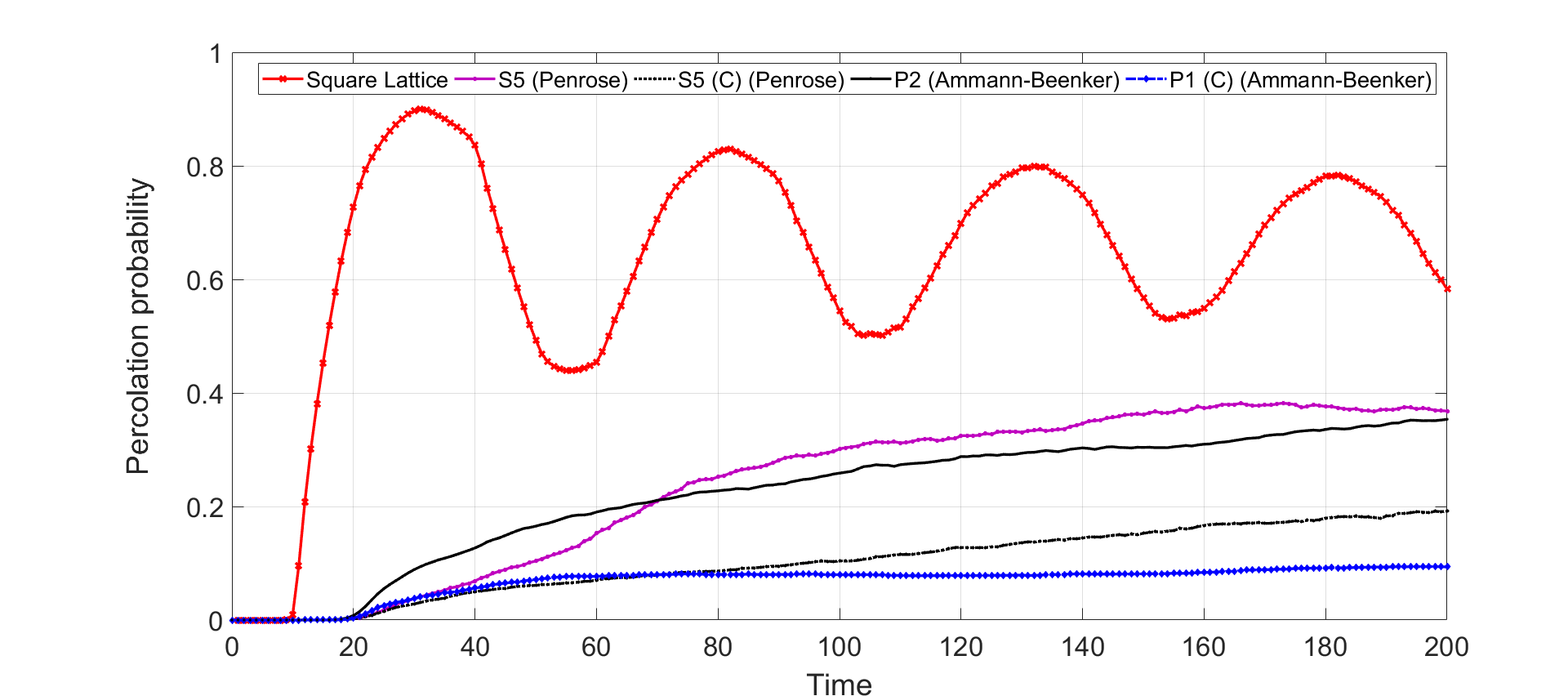} &
			\includegraphics[width=0.5\linewidth]{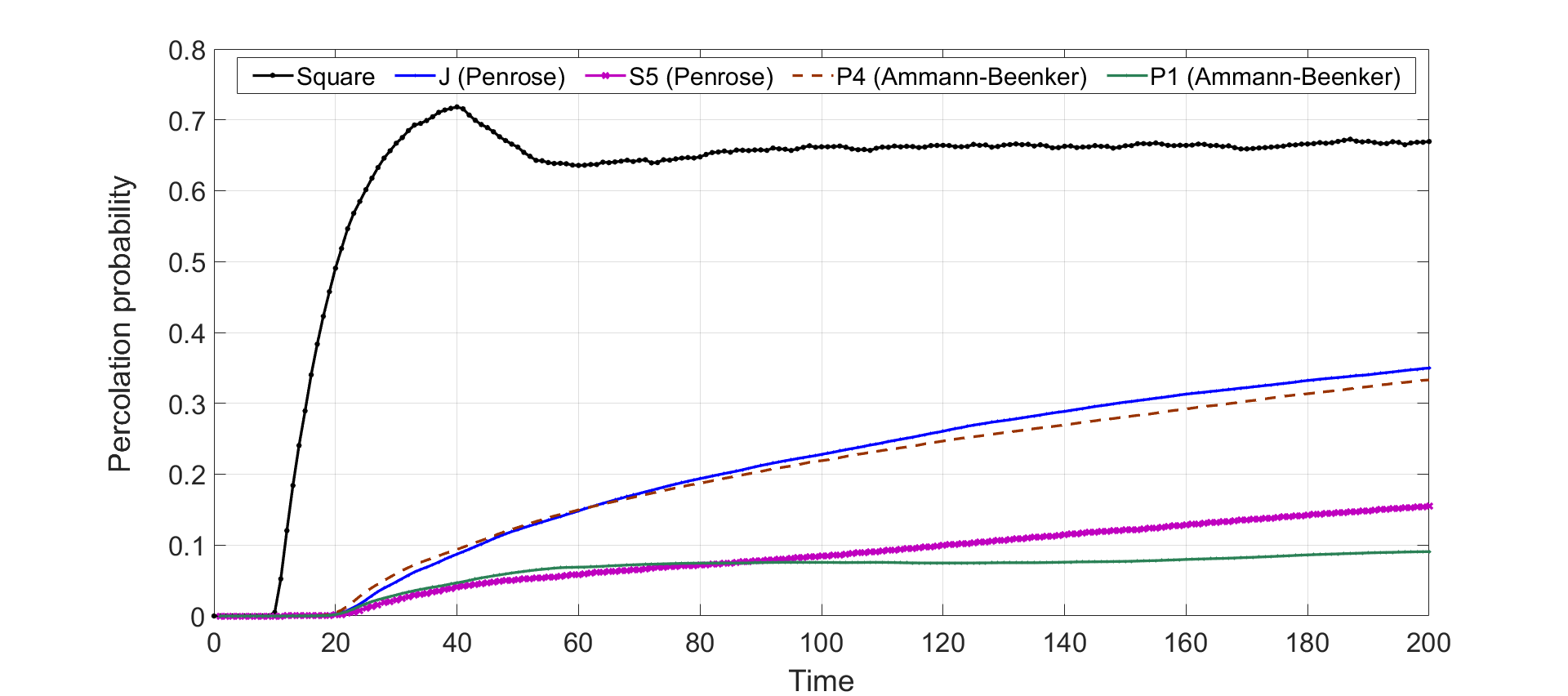} \\
			(a) & (b) \\
			
			\includegraphics[width=0.5\linewidth]{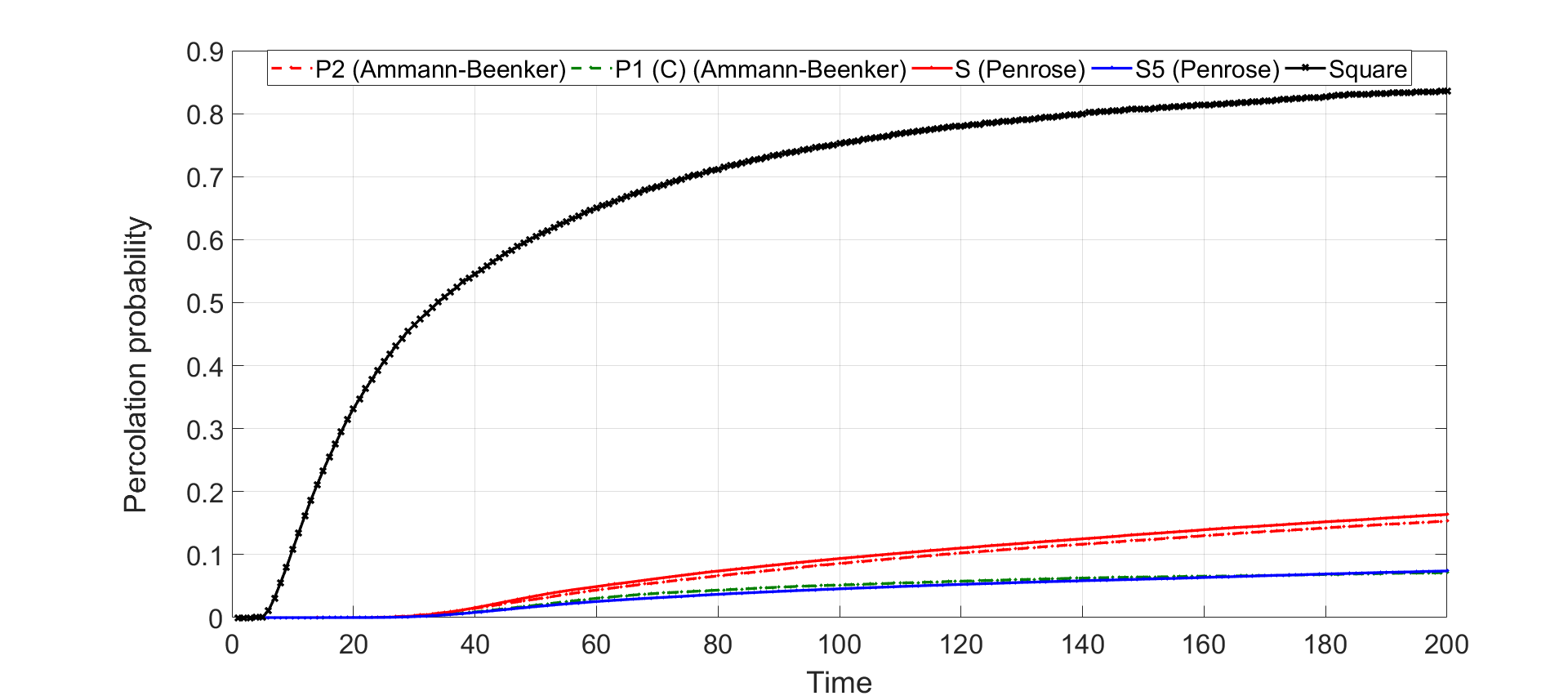} &
			\includegraphics[width=0.5\linewidth]{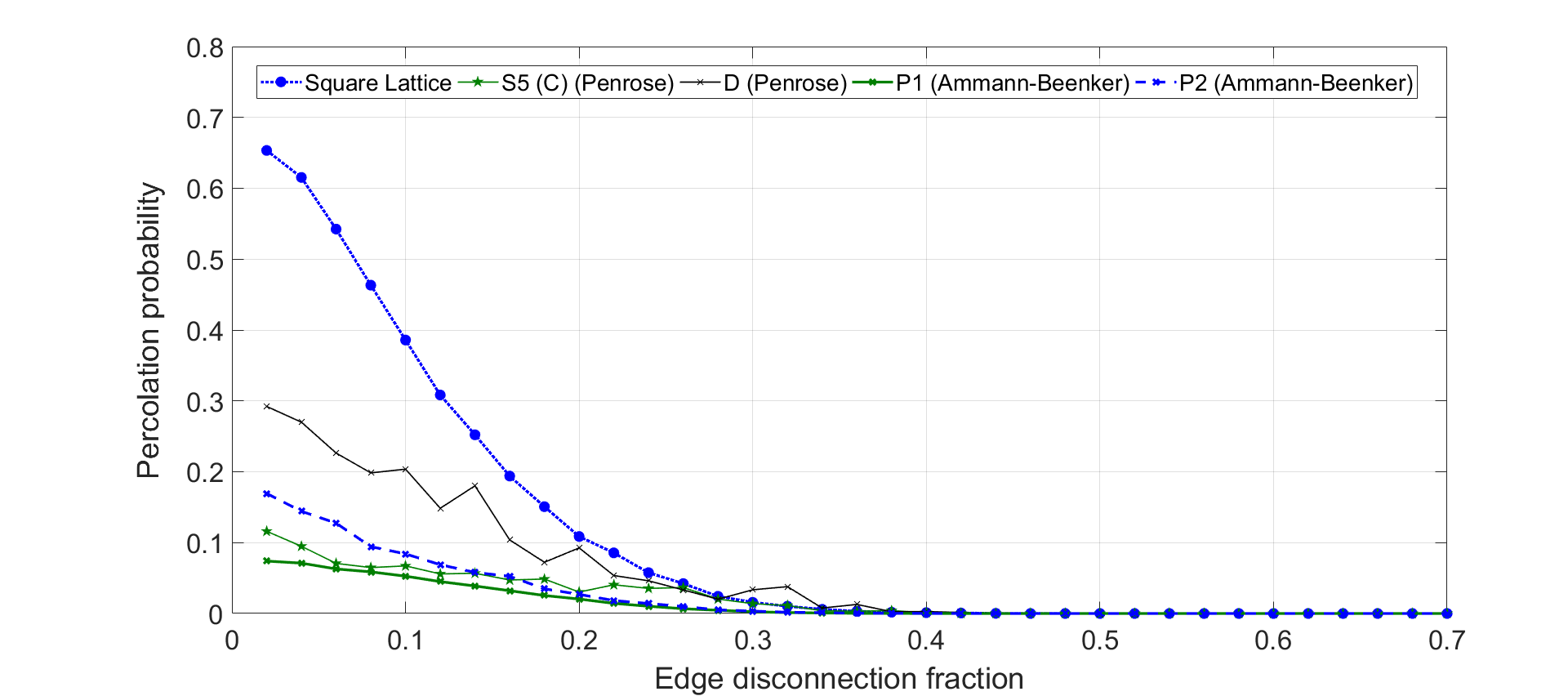} \\
			(c) & (d) 
			
		\end{tabular}
		\captionof{figure}{
			Fig. (a) shows a plot of percolation probability with time for various tilings. The points on the quasicrystal tilings have been chosen to represent the highest and lowest percolation probabilities. \\
			Figs. (b) and (c) show plots of percolation probability with time for various tilings, with 1\% and 10\% disconnected edges, respectively. The points on the quasicrystal tilings were chosen to represent the highest and lowest percolation probabilities. Plots were drawn for a hopping distance 40, and have been averaged over 50 runs.  \\
			Fig. (d) shows a plot of percolation probability with edge disconnection fraction for various tilings, for a hopping zone of length 40 and time 200. The points on the quasicrystal tilings have been chosen to represent the highest and lowest percolation probabilities.
			\label{fig:latticesCombined}
		}
	\end{minipage}
\end{table}

\end{document}